\documentclass{ws-procs975x65standard}
\begin{document}

\title{An updated nuclear reaction network for BBN.}
\author{Pasquale Dario Serpico}
\address{Max-Planck-Institut f\"ur Physik \\(Werner-Heisenberg-Institut)
\\ F\"ohringer
Ring 6 D 80805 M\"unchen\\
\texttt{serpico@mppmu.mpg.de}}

\maketitle \abstracts{The key Standard-Physics inputs of the Big
Bang Nucleosynthesis (BBN) are the light nuclei reaction rates.
Both the network and the nuclear rates have been recently
reanalyzed and updated, and cosmological and New-Physics
constraints (taking into account the WMAP Cosmic Microwave
Background anisotropies measurement) obtained with a new code are
presented.}

Early Universe is a (hot) plasma in a FLRW metric whose
composition and properties depend on the "cosmic" temperature T
and that cools during universe expansion. The main events of its
evolution depend on the \emph{freezing} of some interaction during
the cosmic expansion. BBN takes place when nuclear reactions
(keeping baryons in chemical equilibrium) freeze-out, thus
producing a characteristic pattern in light nuclide abundances.

BBN plays a fundamental role in {\it Cosmology}, where it can be
used to check the internal consistence of the Standard
Cosmological Model (SCM); in {\it Astrophysics}, e.g. to study the
Li depletion mechanism in halo PopII stars, PopIII chemical
composition or the Galactic Chemical Evolution; or to get a hint
of {\it New-Physics}, because BBN is sensitive to the existence of
other relativistic degrees of freedom (parameterized in
$N_{eff}$), to $\nu$'s asymmetries, etc.

In its \emph{minimal formulation}, BBN is an overconstrained
theory: all the relevant observables depend on the only unknown
parameter $\eta\equiv n_b/n_\gamma$, where $n_b$ and $n_\gamma$
are respectively the baryon and the photon number densities. The
other parameters are Standard-Physics inputs, and the greatest
uncertainties in standard BBN predictions arise from nuclear
reaction rates $R_k$.

These rates are obtained as thermal averages of the relevant
cross-sections $\sigma$. From a theoretical point of view, it is
very difficult to use a \emph{first principle approach} (strong
interactions, many body problems, etc.), so one makes recourse to
nuclear \emph{models}. Moreover, experimental difficulties are
present due to the low counting rates, strong energy dependence
and corrections for electron screening. To obtain low energy
extrapolations, fit the data and/or the theoretical predictions,
it is useful to introduce some meaningful parameterizations, as
the so-called {\it astrophysical S factor}.\\{}\\ In a completely
general approach, the error matrix (due to nuclear uncertainties)
for the nuclide abundances is given by the quantities:
\begin{equation}
\sigma_{ij}^2(\theta,R)\equiv \frac 14 \sum_k \left[X_i(\theta,R_k
+ \delta R_k^+)-X_i(\theta,R_k - \delta R_k^-)\right] \times
\left[i\rightarrow j \right]\label{sigmath}
\end{equation}
with $\delta R_k^\pm$ the (temperature dependent) upper and lower
uncertainties on $R_k$, respectively; $R$ represents the
collection of the nuclear reaction rates and $\theta$ the
collection of the other relevant cosmological parameters (i.e.
$\eta, N_{eff},\ldots$). This slightly differs from the approach
in \cite{flsv} which assumes the existence (in principle not
necessary) of the linear functionals $\lambda_{ik}=\partial \log
X_i(\theta) /\partial \log R_k$. To properly define these
quantities, the symmetric, temperature independent, relative
uncertainties $\delta R_k/R_k$ are needed.
\\{}\\ The analysis performed in \cite{mythesis} (see also
\cite{bbnwmap,bbninp}) is restricted to a reduced network
including the reactions relevant for the abundances of the
nuclides with $A \le 7$.\\
More than 80 reactions were examined, and many of the main
reaction rates have been updated. Particularly useful tools in
this work were furnished by the NACRE compilation \cite{nacre},
the LUNA measurement \cite{luna}, or some recent theoretical
predictions (e.g. \cite{rupak}, for the key reaction
$p+n\rightarrow \gamma + {^2H}$). Moreover, in the new code the
nuclear partition functions were introduced to include the role of
excited states for nuclides whose mass number $A \ge 6$
\cite{mythesis,nacre}, and new reactions were added (as the
${^3He}+{^3H}\leftrightarrow \gamma + {^6Li}$, see
\cite{mythesis,bbnwmap,bbninp}). The negligible role of plasma
screening effects and of new three-body reactions was also
confirmed \cite{mythesis,bbninp}.\\{}\\ In the recent paper
\cite{bbnwmap}, some applications of this new code were performed:
we (mainly) checked the internal consistence of the SCM (sections
4 and 5) and evaluated the BBN constraints on some New-Physics
scenarios (section 7): e.g., we showed that in the Degenerate-BBN
scenario a fourth sterile neutrino, as for example required to
interpret LSND evidence for $\overline{\nu}_\mu \leftrightarrow
\overline{\nu}_e$ oscillation, is not yet ruled out.

\end{document}